\documentclass[aip,cha,amsmath,amssymb,reprint,showkeys,floatfix]{revtex4-1}
\usepackage{graphicx,hyperref}

\hypersetup{
       pdfauthor={Tosolini, Anton and Patzauer, Maximilian and Krischer, Katharina},
       pdftitle={Bichaoticity induced by inherent birhythmicity during the oscillatory electrodissolution of silicon},
       pdfsubject={DOI: 10.1063/1.5090118},
       pdfkeywords={Chaos theory, Dynamical systems, Electrochemistry},
       breaklinks=true,
}
\begin{document}
\title[Bichaoticity Induced by Inherent Birhythmicity during the Oscillatory Electrodissolution of Silicon]{Bichaoticity Induced by Inherent Birhythmicity during the Oscillatory Electrodissolution of Silicon}
\author{Anton Tosolini}
\altaffiliation{These authors contributed equally to this work.}
\author{Maximilian Patzauer}
\altaffiliation{These authors contributed equally to this work.}
\author{Katharina Krischer}
\altaffiliation{krischer@tum.de}
\affiliation{
Nonequilibrium Chemical Physics, Department of Physics, Technical University of Munich, 85747 Garching, Germany
}
\begin{abstract}
\label{sec:abstract}
The electrodissolution of p-type silicon in a fluoride-containing electrolyte is a prominent electrochemical oscillator with a still unknown oscillation mechanism. In this article, we present a study of its dynamical states  occurring in a wide range of the applied voltage - external resistance parameter plane. We provide evidence that the system possesses inherent birhythmicity, and thus at least two distinct feedback loops promoting oscillatory behaviour. The two parameter regions in which the different limit cycles exist are separated by a band in which the dynamics exhibit bistability between two branches with different multimode oscillations. Following the states along one path through this bistable region, one observes that each branch undergoes a different transition to chaos,  namely a period-doubling cascade and a quasiperiodic route with a torus-breakdown, respectively, making Si electrodissolution one of the few experimental systems exhibiting bichaoticity.
\end{abstract}
\maketitle
\begin{quotation}
\label{sec:lead}
Bistability is mainly associated with the existence of two stable stationary states at the same set of parameters, a phenomenon  ubiquitous in nonlinear systems. However,  different kinds of attractors, such as a limit cycle and a stationary state, or two limit cycles, might also coexist. The latter type of bistability is commonly referred to as birhythmicity. The coexisting limit cycles often share a common variable in their feedback loops. Hence the limit cycles interact with each other, inducing more complex oscillation modes, such as compound oscillations or even chaos. 
Experimental observations of such dynamics are rare \cite{Alamgir1983, Alamgir1984, Johnson1991}. 
In this article we report bichaoticity in an inherently birhythmic medium, the electrodissolution of silicon in a fluoride-containing electrolyte. 
The coexisting chaotic dynamics emerge through different prominent routes to deterministic chaos coexisting in parameter space. 
\end{quotation}
\section{Introduction}
\label{sec:intro}
The coexistence of multiple attractors is a characteristic feature of many nonlinear dynamical systems. 
A particular type of multistability which has received considerable attention, especially in the biochemical and biological context, is birhythmicity, i.e. the coexistence of two stable limit cycles, each possessing its own characteristic amplitude and frequency \cite{Goldbeter1997}. 
Birhythmicity might arise from the existences of at least two intrinsic feedback loops in a dynamical system. 
An early demonstration of birhythmicity was given by Decroly and Goldbeter \cite{Decroly1982}, who investigated a model comprised of coupled enzymatic reactions with two positive feedback loops in series. 
An experimental system exhibiting birhythmicity was designed by Epstein et al. by linking two chemical oscillators through a common species \cite{Alamgir1983, Alamgir1984}. 
Another early experimental example of birhythmicity includes the exothermic oxidation of hydrogen in a continuously stirred tank reactor \cite{Johnson1991}. 
All these examples have in common that the birhythmic region is confined to a rather small island in parameter space and they all exhibit complex and compound oscillations bearing properties of the two base limit cycles close to the birhythmic parameter range. 
Despite these early findings of intrinsic birhythmic behaviour, experimental studies remain rare, and in particular, the different complex oscillations resulting from the interaction of the two intrinsic oscillatory modes are poorly investigated.

The amount of theoretical studies by far exceeds the number of experimental ones, putting an emphasis on different aspects of birhythmicity, such as spatio-temporal patterns or the important question of how birhythmicity can be controlled. 
For example, Battogtokh and Tyson \cite{Battogtokh2004} discuss conditions that support weak turbulence in birhythmic media, and Stich et al. \cite{Stich2001,Stich2002} demonstrate that self-organised stable pacemakers may exist in reaction-diffusion systems close to the onset of birhythmicity. 
Concerning the control of birhythmicity, Biswas et al. \cite{Biswas2016,Biswas2017} introduced a self-feedback approach capable of eliminating birhythmicity as well as selecting one of the two limit cycles in the birhythmic regime of the system. 
Another approach to control of birhythmicity is time-delay feedback control \cite{Yanchuk2009,Erneux2008,Ghosh2011}.
These works demonstrate that a strong feedback gain as well as large delays can induce bi- as well as multi-rhythmicity. 
This, at first theoretical result, has been experimentally validated with the oscillatory electrodissolution of Cu \cite{Nagy2015}.

In the present study we demonstrate with experiments that the dynamic behaviour due to the interaction of the two oscillatory modes in an inherently birhythmic system can be much more intricate than observed hitherto. 
More precisely, we demonstrate that the electrodissolution of Si in fluoride-containing electrolyte possesses two types of inherent oscillations whose interaction leads to the coexistence of complex oscillation modes as well as to the bistability between a simple limit cycle and complex oscillations. 
This work extends earlier work of our group where we distinguished between two types of oscillations in so-called voltage-jump experiments where the applied voltage was stepped from open circuit to the desired voltage \cite{Miethe2009, Miethe2012, Schoenleber2012}. 
In the present study, we used different protocols for voltage variations which allowed us to access a much larger part of parameter space, and in particular to study the transition between the two types of oscillations in detail. 
Thereby, the coexistence of two prominent routes to deterministic chaos, namely a period doubling cascade and a secondary Hopf bifurcation followed by a torus-breakdown is observed.

The paper is structured as follows. 
In the next section the experimental set-up is described. 
Section \ref{sec:res} starts with a short description of the electrochemical and chemical steps occurring during the anodic dissolution of Si, and is followed by a presentation of typical cyclic voltammograms (CV). 
Then, the two different types of oscillations present in the system are introduced. 
Thereafter an overview of the dynamics in the extended parameter space is given, followed by detailed investigations of certain paths in the parameter plane. 
The results are discussed in section \ref{sec:dis} and we end with conclusions in section \ref{sec:conclusion}.
\section{Experimental system}
\label{sec:exp}
The experimental set-up consisted of a custom made three electrode electrochemical cell. p-type silicon with a (111) crystal orientation and an ohmic resistivity of 5-25\,$\Omega$cm served as a working electrode (WE). 
To obtain an ohmic back contact a 200\,nm thick aluminium layer was thermally evaporated onto the back of the silicon-substrate and subsequently annealed at 400$^\circ$C and $5 \cdot 10^{-5}$\,bar. 
The silicon sample was then mounted on a custom made PTFE holder by contacting it with silver paste and sealing it with red silicone rubber (Scrintex 901, Ralicks GmbH, Rees-Haldern, Germany). 
The freely accessible electrode area was determined using an optical microscope with a precision of approx. 0.1 mm$^2$ and varied between measurement series between 15-25\,mm$^2$. 
The mounted electrode was then cleaned by first gently wiping it with a precision wipe drenched in acetone (Merck, p.a.) and then sequentially immersing it in acetone (Merck, p.a.) (10\,min), ethanol (Merck, p.a.) (5\,min), methanol (Merck, p.a.) (5\,min) and finally in ultra pure water (18.2\,M$\Omega$cm) (5\,min).
The electrolyte contained 142\,mM H$_2$SO$_4$ (Merck, Suprapur) and 0.06\,M NH$_4$F (Merck, p.a.) resulting in a pH value of 1 according to dissociation constants found in the literature \cite{Cattarin2000}. 
A circularly shaped platinum wire served as a counter electrode and a saturated mercury/mercury-sulphate electrode served as a reference electrode. Note, however, that below all voltages are given with respect to the standard hydrogen electrode (SHE). 
The potential was controlled with a potentiostat, either FHI-2740 for measurements where the applied potential was varied coarsely or Biologic SP-200 for measurements where the applied potential was varied finely. Additionally, an adjustable ohmic resistance was placed in series to the WE and served as a control parameter.

The electrochemical experiments were conducted in combination with ellipsometric imaging to characterise the oxide layer on the WE. 
The optical set-up is the same as implemented by Miethe et al. \cite{Miethe2012}. 
The measured signal is a light intensity of an optical beam reflected from the electrode surface and is measured using a CCD (JAI CV-A50).
The flux of photons onto the CCD gives a linear response until a threshold is reached and the signal saturates.
Our measured signal will be given relative to this saturation level and is from here on denoted as ellipsometric intensity. 
In accordance with earlier measurements with p-type Si \cite{Miethe2012, Schoenleber2016}, oscillations in the ellipsometric intensity signal were spatially uniform in the entire parameter region investigated here. 
Therefore, only the spatially averaged ellipsometric intensity signal $\xi$ is presented. 
The oscillation amplitude of $\xi$ is small compared to its absolute value and thus shows an approximately linear dependence on the optical path length through the oxide layer during the oscillations \cite{Miethe2012}.
\section{Results}
\label{sec:res}
\subsection{Cyclic Voltammetry and Oscillation Types}
\label{sec:CV}
\begin{figure}[tbp!]
	\centering
	\includegraphics[width=0.48\textwidth]{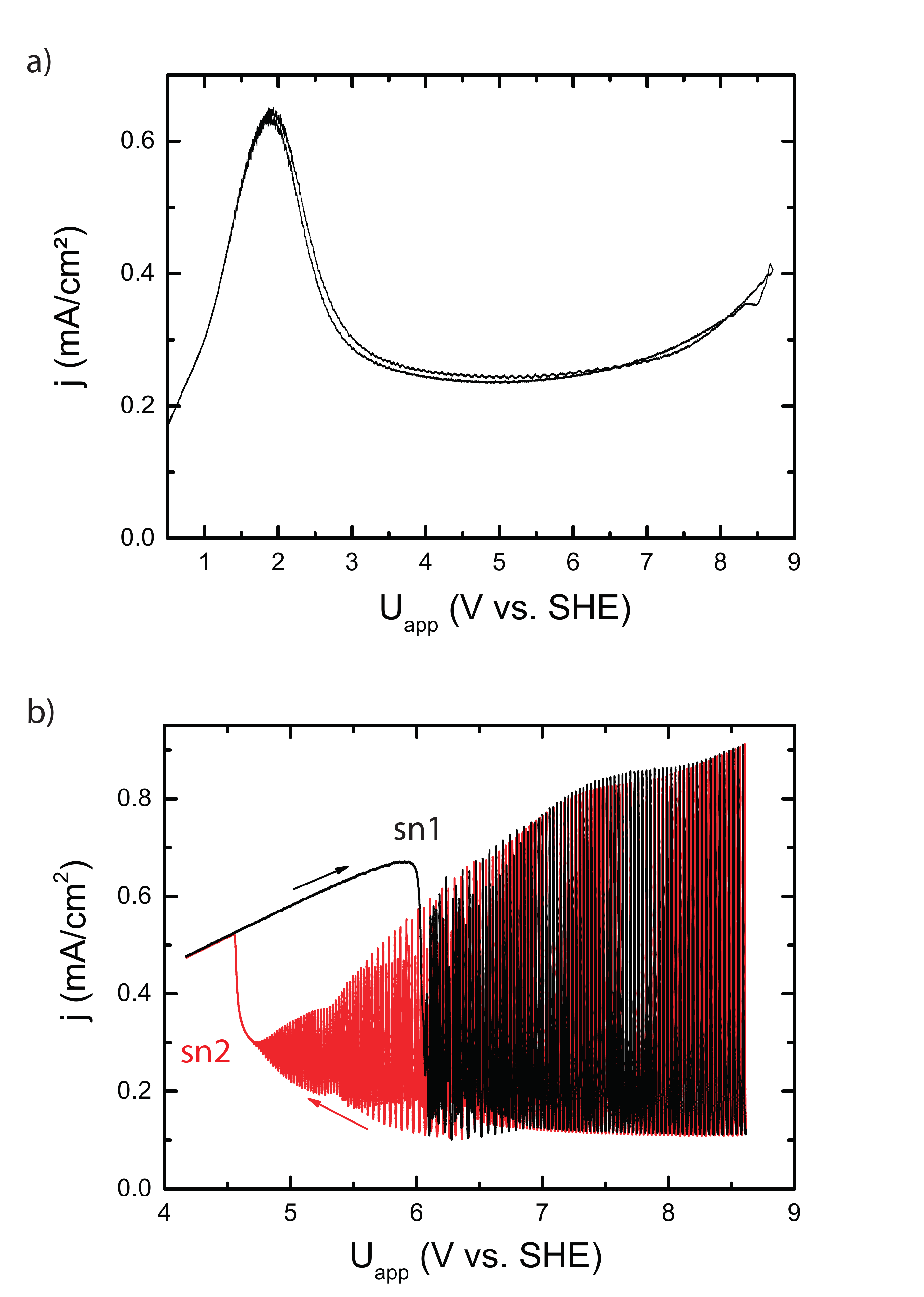}
	\caption{CVs of a p-type Si electrode in 0.06\,M $\text{NH}_4\text{F}$ at pH 1 a) without and b) with an applied external resistance ($R_\text{ext}A=6$\,k$\Omega$cm$^2$), black - forward scan, red - backward scan. Scan rates: a) 2\,mV/s, b) 1\,mV/s.}
	\label{fig:cvwrapp}
\end{figure} 

\begin{figure}[tbp!]
	\centering
	\includegraphics[width=0.48\textwidth]{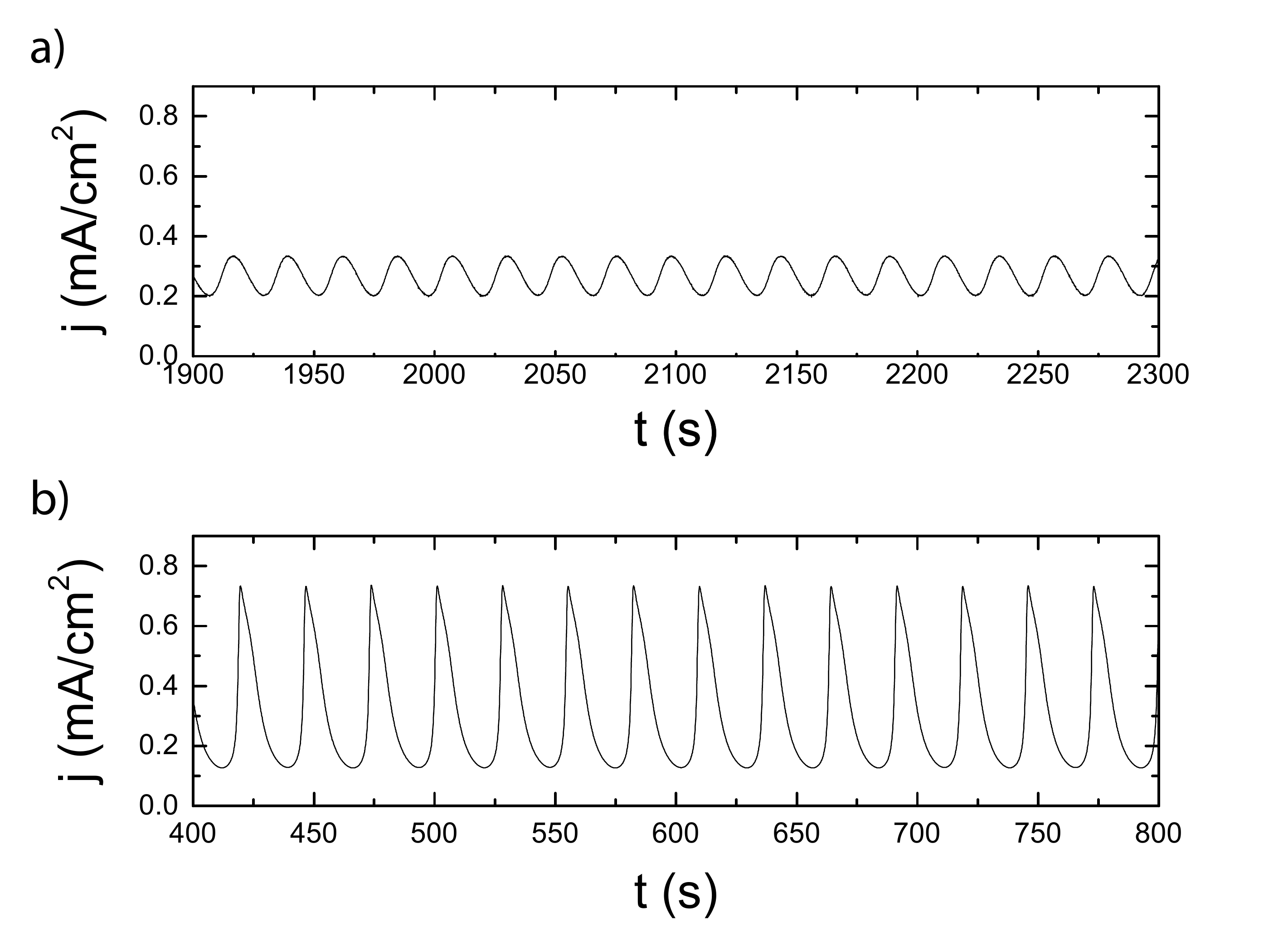}
	\caption{Time series of a) LA oscillations measured at 4.98\,V~vs.~SHE and b) HA oscillations at 6.85\,V~vs.~SHE. Both states measured at $R_\text{ext}A=6$\,k$\Omega$cm$^2$.}
	\label{fig:HAts}
\end{figure}

Fig.\,\ref{fig:cvwrapp}\,a) depicts a typical CV of p-type Si in a fluoride-containing electrolyte at pH 1 with a scan rate of 2\,mV/s. 
Its shape results from the competition between the electrochemical oxidation of the Si electrode to silicon oxide and the chemical etching of the oxide. 
The CV can be roughly divided in three regions: 
(i) the electropolishing branch below about 2\,V where the current density increases strongly and the chemical etching of Si oxide is faster than its electrochemical oxidation \cite{Eddowes1990,Blackwood1992}, 
(ii) the region of negative differential resistance (NDR) between about 2\,V and 3-4\,V where a so called 'wet', non-stoichiometric SiO$_x$ layer builds up on the electrode surface \cite{Salman2019}, and (iii) the plateau region above 3 to 4 \,V where the oxide layer consists dominantly of stoichiometric SiO$_2$. 
It is in region (iii) where oscillations are observed. For an overview of early work on oscillations in this system see chapter 5 in Ref. \cite{Zhang2001}.
If no external resistance is inserted in series to the WE, as in the CV of Fig.\,\ref{fig:cvwrapp}\,a), the current oscillations are tiny \cite{Patzauer2017}. They become pronounced in the presence of an external resistance in series to the WE, as can be seen in Fig.\,\ref{fig:cvwrapp}\,b) where a CV with a comparatively large external resistance is shown. 
In Fig.\,\ref{fig:cvwrapp}\,b) it can be seen that the external resistance tilts the CV towards larger values of the applied potential $U_\text{app}$ such that the steady state on the electropolishing branch is annihilated in a saddle node bifurcation and the system directly enters the oscillatory region from the maximum of the electropolishing branch. 
On the backward scan the system stays in the oscillatory state down to a second sn-bifurcation at lower potentials, thereby describing a hysteresis. 
The occurrence of two sn-bifurcations enclosing a bistable regime is a well known behaviour for N-shaped NDR-systems under load \cite{Koper1992}.
In Fig.\,\ref{fig:cvwrapp}\,b) it is striking that the oscillation amplitude changes drastically with voltage: 
at low values of $U_\text{app}$ the oscillations have a low amplitude; at high values of $U_\text{app}$ their amplitude increases to values higher than the current maximum on the electropolishing branch. 
Furthermore, they seem to be simple periodic at low and high $U_\text{app}$ values but take on complex shapes at intermediate values. 
In fact, these characteristics are also found for constant parameter values, 
i.e. when $U_\text{app}$ is kept at a fixed value \cite{Miethe2012, Schoenleber2012, Schoenleber2016}. Examples of time series of "Low Amplitude" (LA) and "High amplitude" (HA) oscillations measured at low and high voltage, respectively, are shown in Fig.\,\ref{fig:HAts}. 
The frequency of the respective oscillations varies depending on the system parameters.
Owing to different electrochemical fingerprints, it had originally been suggested that they  involve inherently different electrochemical processes \cite{Miethe2012}, though later the interpretation was favoured that they emerge from the same mechanism \cite{Schoenleber2012,Blaffart2014}. 
In the following we demonstrate that LA oscillations and HA oscillations indeed live on different branches, and that the transition between both states occurs through complex oscillatory behaviour.
\subsection{Parameter Space}
\label{sec:pspace}
\begin{figure}[tbp!]
	\centering
	\includegraphics[width=0.48\textwidth]{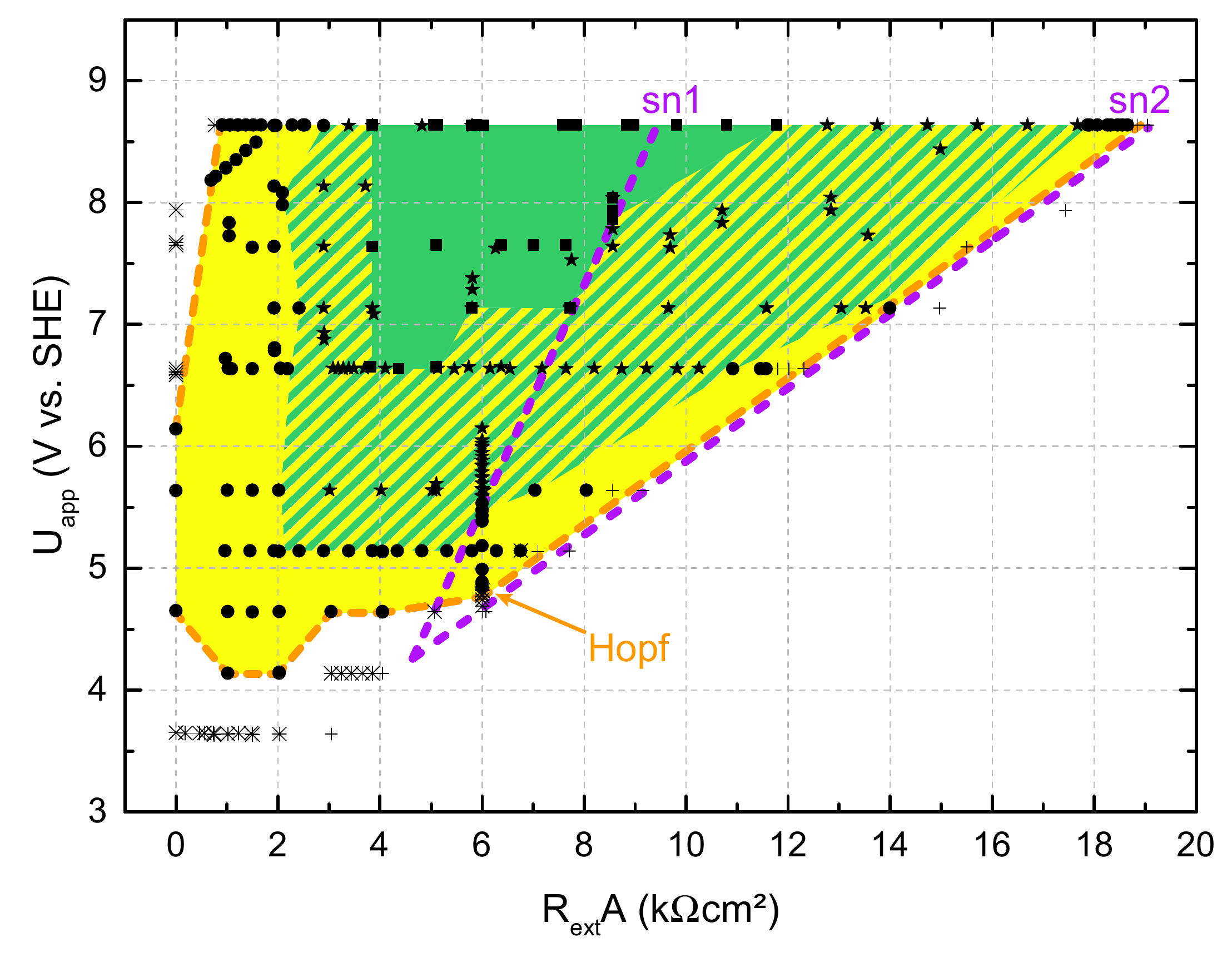}
	\caption{Investigated parameter space with the different types of oscillations, marked
		with: (\textbullet) simple periodic Low Amplitude oscillations (LA), ($\blacksquare$) simple periodic High Amplitude oscillations (HA), ($\bigstar$) multiperiodic and aperiodic oscillations, (\textasteriskcentered) foci and ($+$) nodes. The oscillatory regime of the parameter space is roughly divided into three different regions which are coloured;  yellow LA, green HA, yellow and green striped multiperiodic and aperiodic  oscillations. The dashed orange line marks the Hopf bifurcation from which the LA oscillations arise. The sn1 line represents the border between the sets of parameters for which no initial oxide layer on the electrode is necessary for the system to attain a stable limit cycle (left) and those for which an initial oxide layer is needed (right). The sn2 line marks the transition from the oscillatory regime to the electropolishing branch.}
	\label{fig:pasp}
\end{figure}

Fig.\,\ref{fig:pasp} gives an overview of the dynamical states observed in the system as a function of two bifurcation parameters, $U_\text{{app}}$ and $R_\text{{ext}}A$. 
The coloured area marks the oscillatory region. 
LA oscillations are found in the yellow region and HA oscillations are found in the green region. 
These two regions are separated by the green and yellow striped parameter regime in which a multitude of complex and chaotic oscillations were observed. 
Note, that in some parts of the green region, HA oscillations and complex oscillations coexist. 
Parameter values at which measurements were conducted are indicated by symbols; circles, squares and stars stand for LA oscillations, HA oscillations and complex oscillations, respectively.  
Except for a small $U_\text{app}$ interval between approximately 4.5 and 6.5\,V, where the oscillations extend down to zero external resistance, and thus to the border of the physical sensible $R_\text{ext}$ values, the oscillatory region is enclosed by a Hopf bifurcation. 
The Hopf bifurcation is sketched by the orange dashed lines. 
The two purple lines, sn1 and sn2, mark the two aforementioned sn-bifurcations, see Fig.\,\ref{fig:cvwrapp}\,b). 
They enclose a multistable regime of the parameter space in which the system can attain a stable steady state on the electropolishing branch or an oscillatory state where the Si surface is covered by an oxide layer. 
It is worth mentioning, that the oscillations coexisting with the electropolishing branch, i.e. in the parameter region between the two sn-bifurcations, can only be reached from an initially oxide covered state. 
So far, in the literature the oscillations were almost always established by stepping the potential from open circuit potential where the Si electrode is H-terminated, to large potential values. 
Hence, hitherto oscillations in this parameter region have been missed. 
\subsection{Hopf bifurcation}
\label{sec:hopf}
\begin{figure}[tbp!]
	\centering
	\includegraphics[width=0.48\textwidth]{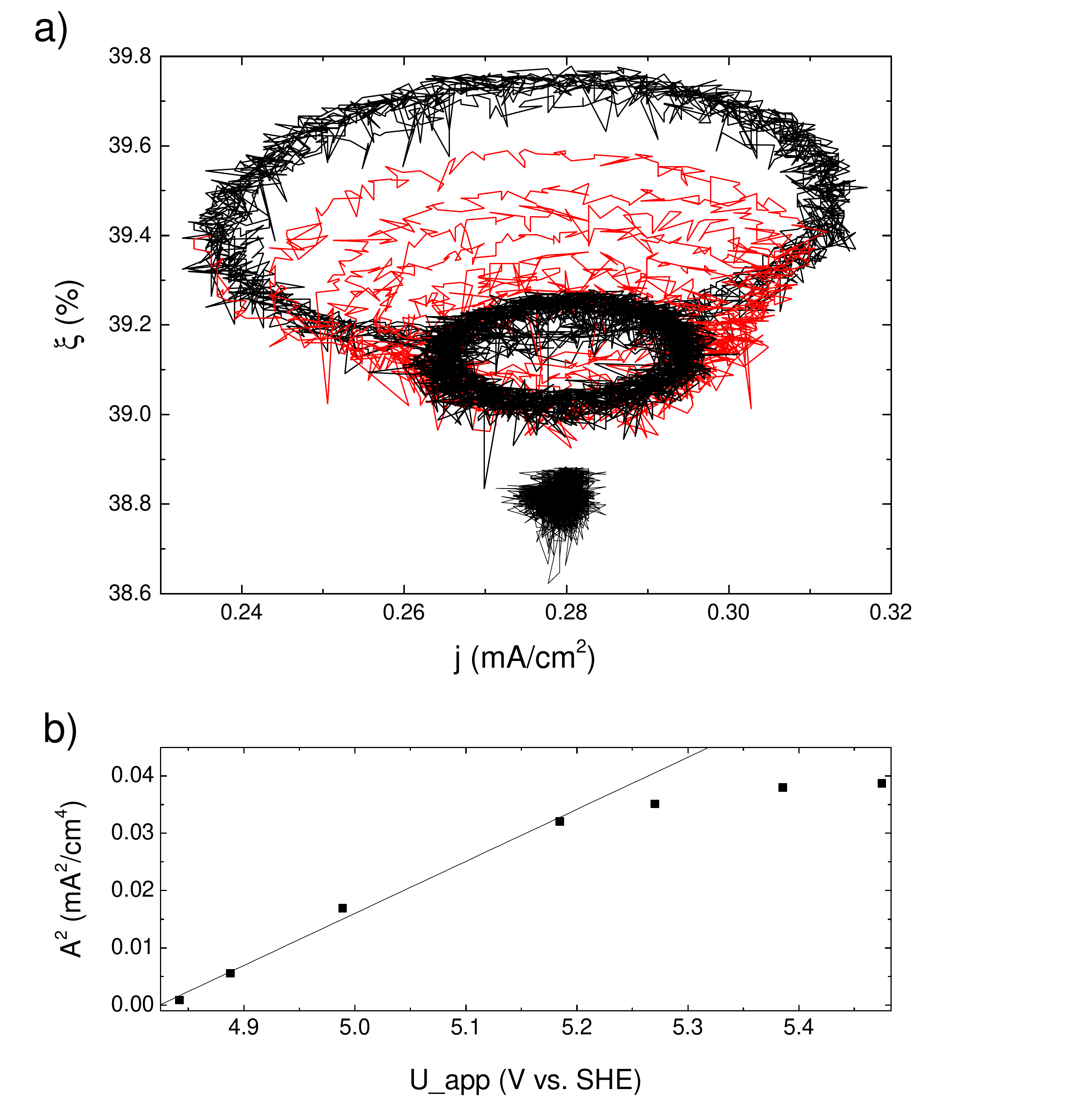}
	\caption{a) LA oscillations and stable focus in the ellipsometric intensity $\xi$ vs. current density j phase plane projection at $R_\text{ext}A=6$\,k$\Omega$cm$^2$. Black lines mark attractors, the red line shows the transient after a parameter variation. The large limit cycle was measured at 4.90\,V~vs.~SHE, the small limit cycle was measured at 4.84\,V~vs.~SHE and the stable focus was measured at 4.82\,V~vs.~SHE.
	b) The squared peak-to-peak amplitude of LA oscillations close to the Hopf bifurcation $A^2$ vs. the applied potential $U_\text{app}$ at $R_\text{ext}A=6$\,k$\Omega$cm$^2$. The line is the linear fit for the first four points.}
	\label{fig:phspinter}
\end{figure}

In order to further elucidate the dynamics in the bistable region close to the sn2-bifurcation, a finely resolved parameter-scan was conducted.
Fig.\,\ref{fig:phspinter}\,a)  shows projections of phase space portraits of potentiostatic measurements with a constant load of $R_\text{ext}A~=~6$\,k$\Omega$cm$^2$ in the ellipsometric intensity $\xi$ vs. current density $j$ plane.

As was mentioned above, to reach states coexisting with the electropolishing state, the initial condition had to be an oxide covered state. 
This was achieved by first letting the system set in an oscillatory state at a higher applied voltage (6.90\,V, $R_\text{ext}A=6$\,k$\Omega$cm$^2$)  and then subsequently changing the potential to 4.90\,V, thereby keeping the applied external resistance constant. 
From there on we waited for the system to settle, recorded the time series and then changed the applied potential to a lower value. 
The black lines in Fig.\,\ref{fig:phspinter}\,a) mark attractors of the system, and the red line indicates a transient after parameter variation. 
As can be seen, at the largest of the depicted $U_\text{app}$ values, a stable LA oscillation exists. 
The limit cycle shrinks with decreasing applied potential until it disappears altogether and the system attains a stable focus at 4.82\,V. In Fig.\,\ref{fig:phspinter}\,b) the squared peak-to-peak amplitude of the current oscillations is plotted against the applied potential. 
The good linear fit through the first four points, strongly suggests that the limit cycle emerges from a Hopf bifurcation. 
Another small variation of the parameters either towards lower applied potential or towards higher resistance drove the system across the sn-bifurcation marked by sn2 in Fig.\,\ref{fig:pasp} to a steady state on the electropolishing branch. 
The Hopf bifurcation and the sn2 were found to lie very close together in the $U_\text{app}$-$R_\text{ext}A$ parameter plane, except very close to the cusp bifurcation, cf. Fig.\,\ref{fig:pasp}. 
Fig.\,\ref{fig:phspinter}\,a) also reveals that the LA oscillations cannot be described in terms of $\xi$ and $j$ alone, since the transient trajectory intersects itself when projected onto the $\xi-j$ phase space plane.
\subsection{Bichaoticity}
\begin{figure}[bp!]
	\centering
	\includegraphics[width=0.48\textwidth]{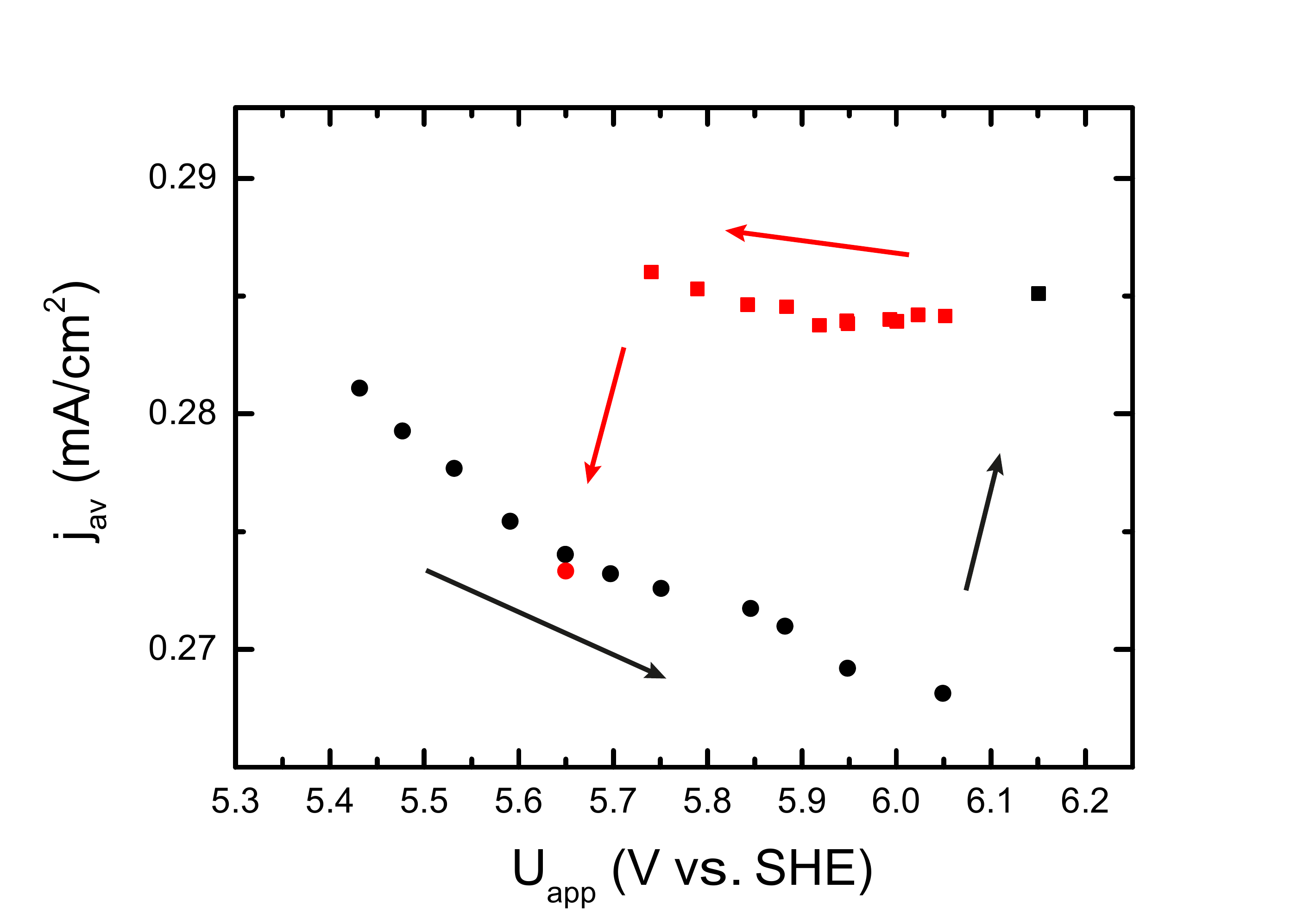}
	\caption{Average current density of measured time series vs. applied voltage. The black points are measured on the way to higher potentials and the red points on the way to lower potentials. The black and red arrows also indicate the order of the measurements. (\textbullet) mark states on the LA branch, ($\blacksquare$) mark states on the HA branch.}
	\label{fig:bist}
\end{figure}

In the following we investigate the transition between LA oscillations and HA oscillations at a constant value of $R_{\text{ext}}A=
6\,\text{k}\Omega$cm$^2$ with $U_\text{app}$ as bifurcation parameter. This path is representative for the striped region in parameter space which shows a multitude of different dynamics. In order to capture transitions between different attractors correctly, the applied potential was varied gently by sweeping it with 1\,mV/s between the measured states. 
Furthermore, we assured that the system had sufficient time to relax to the new attractor before recording the dynamics.
Whenever, we found the system to undergo a qualitative change in its dynamics, we reversed the direction of the potential variations to test for hysteresis. 
Thereby two branches of qualitatively different dynamics could be identified. 
These two branches are visualised in  Fig.\,\ref{fig:bist} where the average current densities of the respective oscillations are plotted against the applied potentials. Black points were measured on the way to higher $U_\text{app}$ values and red points were measured after $U_\text{app}$ was decreased. Clearly the measurements exhibit bistability. In the following, we discuss the dynamics occurring along these two branches in detail.

\begin{figure*}[tbp!]
	\centering
	\includegraphics[width=\textwidth]{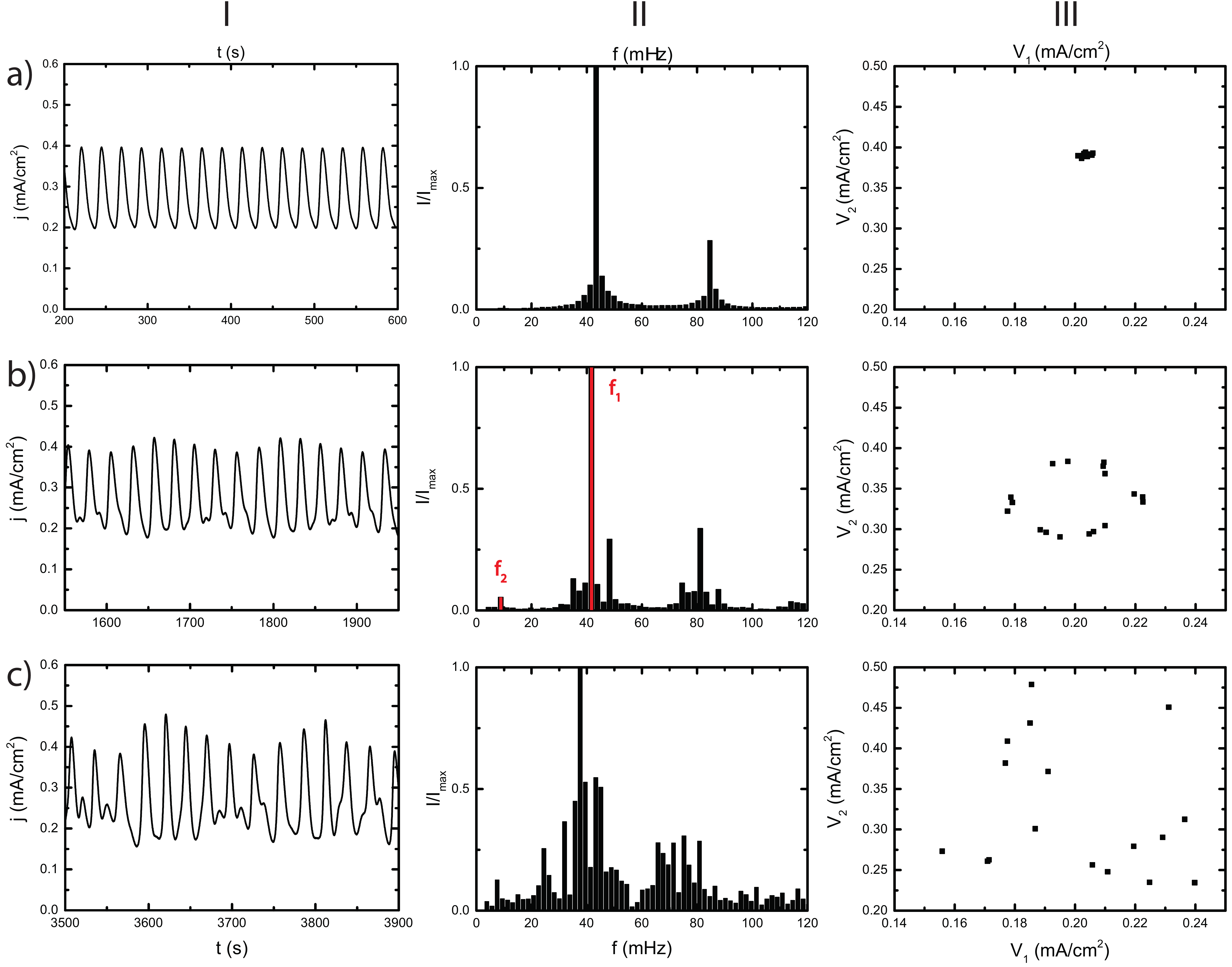}
	\caption{(I) Time series, (II) Fourier spectrum and (III) Poincar\'e section of three  measurements at different voltages; a) 5.54\, V vs.  SHE, b) 5.69\,V vs.  SHE and c) 5.95\,V vs.  SHE, all with an external resistance of $R_\text{ext}A=6$\,k$\Omega$cm$^2$}
	\label{fig:table}
\end{figure*}

Let us first focus on the lower branch of Fig.\,\ref{fig:bist}. Time series, corresponding Fourier spectra and Poincar\'e sections of current measurements at three increasing values of $U_\text{app}$ are depicted in Fig.\,\ref{fig:table}. 

The Poincar\'e sections were constructed by first choosing a suitable current density value $j_0$ and recording the times $t_k$ at which the current times series go through  this value on their falling flank, i.e. the $t_k$ for which $ j(t=t_k) = j_0 $. Subsequently, we introduced time delays to define the variables $V_1$ and $V_2$ such that:
\begin{equation}
V_1(t_{k})=j(t_{k}-3T/4) \text{ and }   V_2(t_{k})=j(t_{k}-T/4).
\label{eq:poincare}
\end{equation}

\begin{figure}[tbp!]
	\centering
	\includegraphics[width=0.48\textwidth]{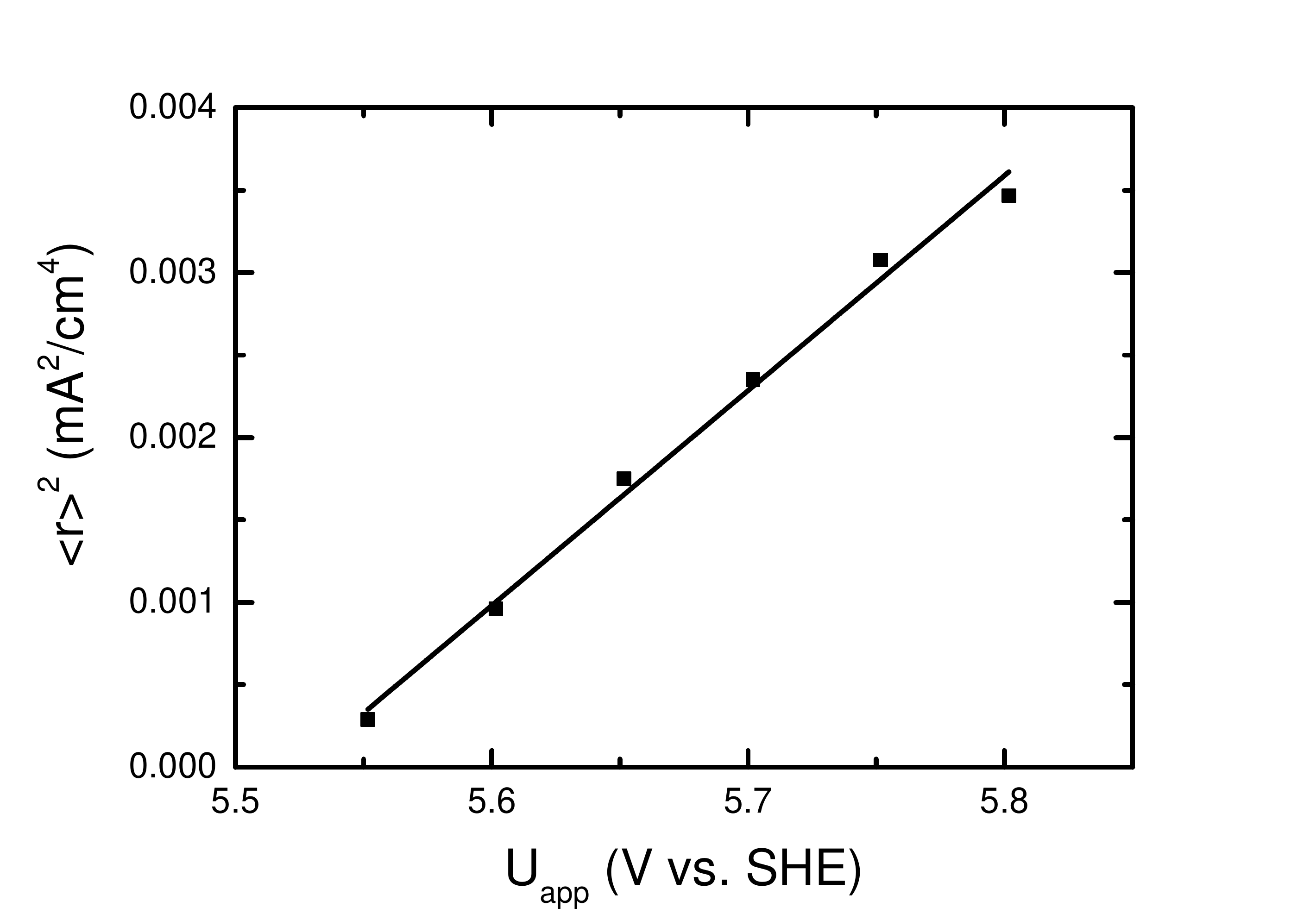}
	\caption{The squared average radius of the invariant circle in the Poincar\'e section as defined in Eq.~\eqref{eq:avnmm} vs.  the applied potential $U_\text{app}$.}
	\label{fig:2ndary}
\end{figure}

Here, $T$ is the period corresponding to the main frequency of the respective Fourier spectrum. 
Fig.\,\ref{fig:table}\,a)\,I shows a simple periodic time series which corresponds to LA oscillations and was obtained at an applied potential of 5.54\,V. In accordance with this simple limit cycle dynamics, the appendant Fourier spectrum (Fig.\,\ref{fig:table}\,a)\,II) is characterised by one main frequency and its first harmonic. 
In the Poincar\'e section (Fig.\,\ref{fig:table}\,a)\,III) the dynamic is mapped to a (somewhat noisy) fixed point. 
Fig.\,\ref{fig:table}\,b)\,I shows the time series after the potential was increased to 5.69\,V. Now, the time series seems to have a beat-like character. 
This interpretation is supported by the Fourier spectrum (Fig.\,\ref{fig:table}\,b)\,I) which now features the main frequency f$_1$ and its harmonic, as well as a second frequency f$_2$, and linear combinations between the frequencies f$_1$ and f$_2$, as is characteristic for quasiperiodic behaviour. 
Furthermore, the dynamics in the Poincar\'e section (Fig.\,\ref{fig:table}\,b)\,III) seems to live on an invariant circle, again in line with our interpretation that at this value of $U_\text{app}$ the dynamics is quasiperiodic and lives on a torus. 
As $U_\text{app}$ is increased further, the radius of the invariant circle in the Poincar\'e section increases. 
In order to determine how the radius of the invariant circle changes as a function of the applied voltage, we determined the averaged radius of the invariant circle in the Poincar\'e section according to Eq.\,\eqref{eq:avnmm},
\begin{equation}
\langle\parallel\vec{x_\text{i}}-\langle\vec{x}\rangle\parallel\rangle^2=\langle r \rangle^2\,,
\label{eq:avnmm}
\end{equation}

with $\vec{x}=(V_1,V_2)^\text{t}$ and triangular brackets denoting the average of all points $\vec{x_\text{i}}$. 
Fig.\,\ref{fig:2ndary} shows a plot of $\langle r\rangle^2$ vs. the applied potential $U_\text{app}$. 
It shows a linear behaviour, which indicates that the simple periodic LA oscillations have undergone a secondary Hopf bifurcation resulting in quasiperiodic motion on a torus as $U_\text{app}$ was increased. 
Fig.\,\ref{fig:table}\,c)\,III shows the Poincar\'e section after the potential was further increased to 5.95\,V. 
Here, clearly the torus has fallen apart, the iterates of the dynamics being irregularly scattered in the Poincar\'e section. 
Furthermore, in the corresponding Fourier spectrum (Fig.\,\ref{fig:table}\,c)\,II) the background is significantly increased. 
These observations suggest that the dynamics is chaotic. We have not found any indications for a tertiary Hopf bifurcation. 
Although this does not necessarily exclude the Ruelle-Takens-Newhouse scenario as a route to chaotic motion, we consider it more likely that the transition to chaos occurs through the loss of smoothness of the T$^2$ torus, as for example described in Ref.  \cite{Afraimovich1991,Anishchenko1993}.

\begin{figure}[bp!]
	\centering
	\includegraphics[width=0.48\textwidth]{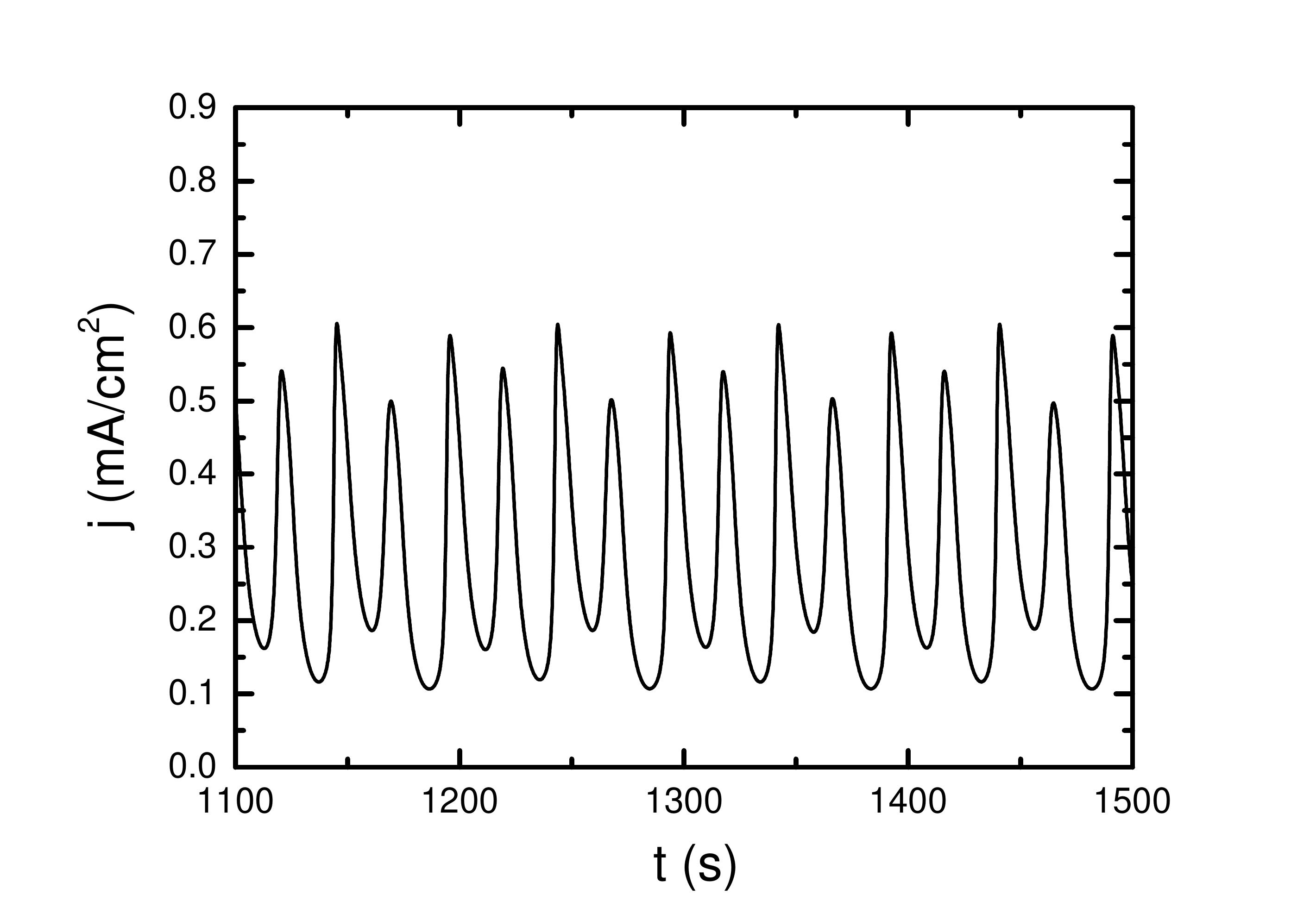}
	\caption{Time series of a period-4 state on the HA branch measured at 6.15\,V vs.  SHE.}
	\label{fig:P4ts}
\end{figure}

\begin{figure}[bp!]
	\centering
	\includegraphics[width=0.48\textwidth]{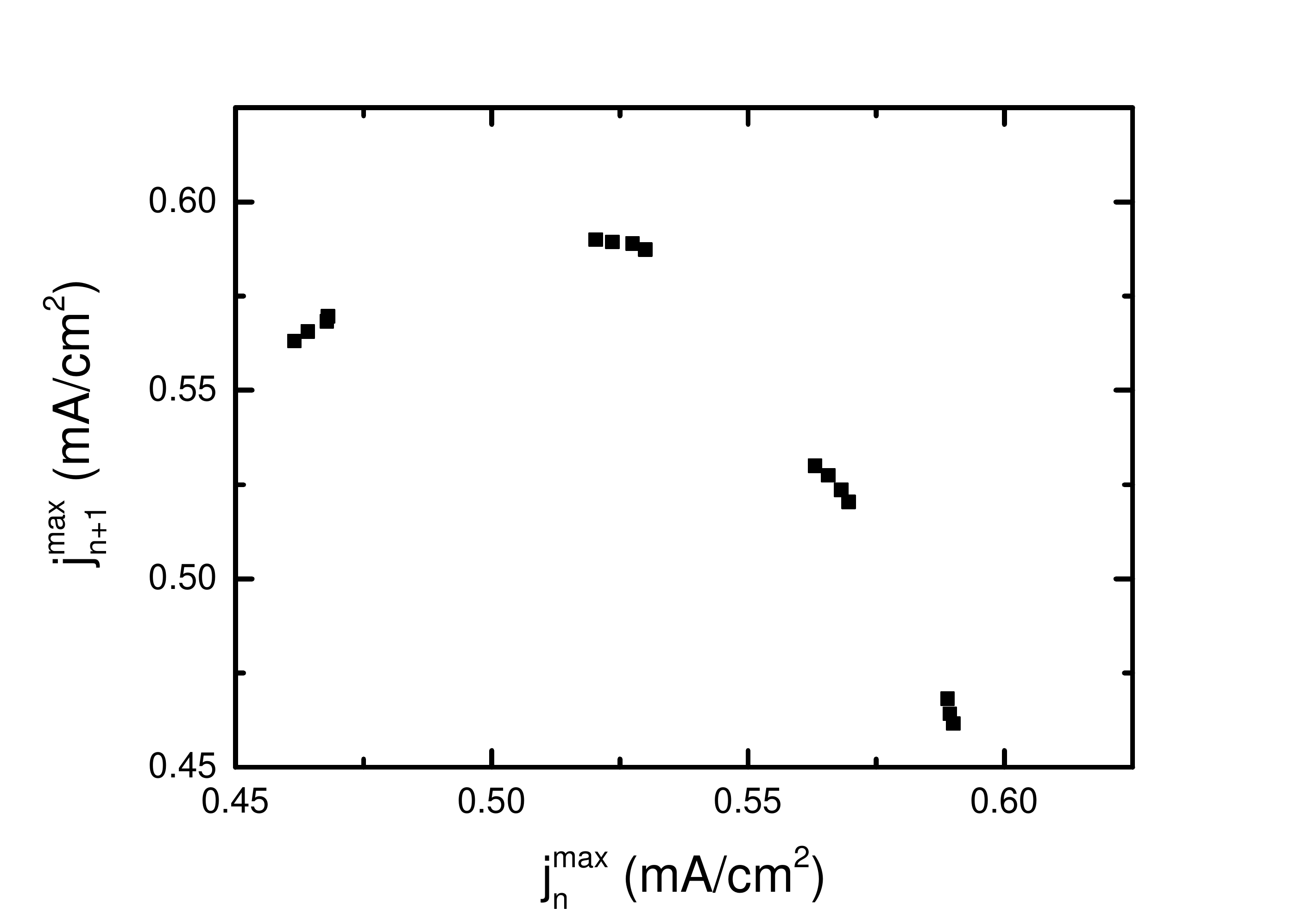}
	\caption{Next-maximum-map of a chaotic four-band attractor measured at 6.05\,V vs.  SHE.}
	\label{fig:C4nmm}
\end{figure}

Starting from this chaotic motion, an increase of the applied potential eventually leads to an abrupt change of the dynamics. 
Fig.\,\ref{fig:P4ts} shows the corresponding time series. 
\newline  \indent 
The amplitude of these oscillations is significantly larger than of those on the lower branch and the oscillations now have a multi-mode character. 
Moreover, the average current density is considerably larger than for all the states before, as can be seen in Fig.\,\ref{fig:bist}. 
When decreasing $U_\text{app}$ again, in fact we enter a parameter interval in which the dynamics is bistable, whereas when
increasing $U_\text{app}$ further the higher periodic oscillations of Fig.\,\ref{fig:P4ts} bifurcate into simple periodic HA oscillations. 
The lower one of the two branches in Fig.\,\ref{fig:bist} is thus connected with LA oscillations, the upper one with HA oscillations.
\begin{figure}[tbp!]
	\centering
	\includegraphics[width=0.48\textwidth]{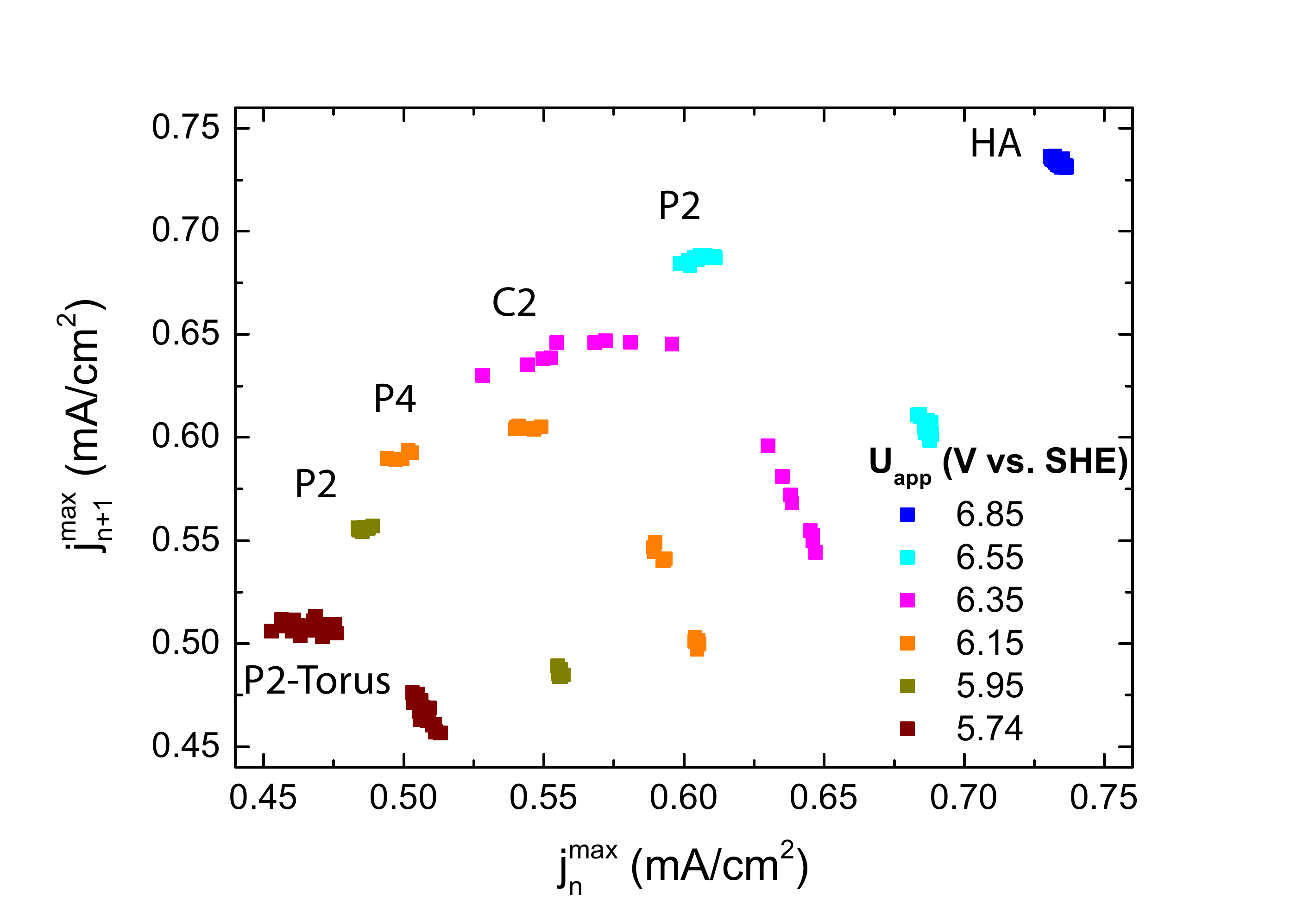}
	\caption{Next-maximum-maps of the states measured on the HA branch. Different colours are measurements at different applied potentials as indicated in the inset. HA: simple periodic HA oscillation; P2: period-two oscillation; P4: period-four oscillation; C2: chaotic two band attractor; P2-Torus: quasiperiodic motion on a period-2 torus.}
	\label{fig:feignmm}
\end{figure}
\newline \indent
To obtain more insight into the dynamics on the upper branch, the next-maximum-maps of time series at various voltages are depicted in Fig.\,\ref{fig:feignmm}. 
It suggests that, when coming from high applied voltages, HA oscillations become unstable via a period doubling bifurcation, which seems to be part of a period doubling cascade to chaos. 
At $U_\text{app}=6.35$\,V the dynamics lives on a chaotic two-band attractor, denoted in Fig.\,\ref{fig:feignmm} as C2. 
Upon further decreasing $U_\text{app}$, a period halving cascade is observed up to the state with a period-2 dynamics, P2, at $U_\text{app}=5.95$\,V. 
The system thus exhibits bistability between two qualitatively different routes to chaos, a torus breakdown and a Feigenbaum scenario. 
\newline \indent
To demonstrate the degree of reproducibility of the experiment, we show an example of a next-maximum-map of a chaotic four-band attractor in Fig.\,\ref{fig:C4nmm} which was recorded within another measurement series at $U_\text{app}=6.05$\,V. 
\newline  \indent This parameter value lies between the ones at which the period-2 and period-4 oscillations were observed in the measurement series depicted in Fig.\,\ref{fig:feignmm}. 
In contrast, the Feigenbaum scenario predicts that the C4 band attractor exists at a parameter value between the ones at which P4 and C2 dynamics are found. 
This type of slight shift of the parameter values at which bifurcations were observed was typically found when comparing different measurement series, but the qualitative sequence of the bifurcations always remained the same. 
We attribute this quantitative shift of the bifurcation parameters to slight variations of the experimental parameters, which were not controlled during the experiments.This includes for example temperature and electrolyte composition, which changes due to the consumption of fluoride.

\begin{figure}[tbp!]
	\centering
	\includegraphics[width=0.48\textwidth]{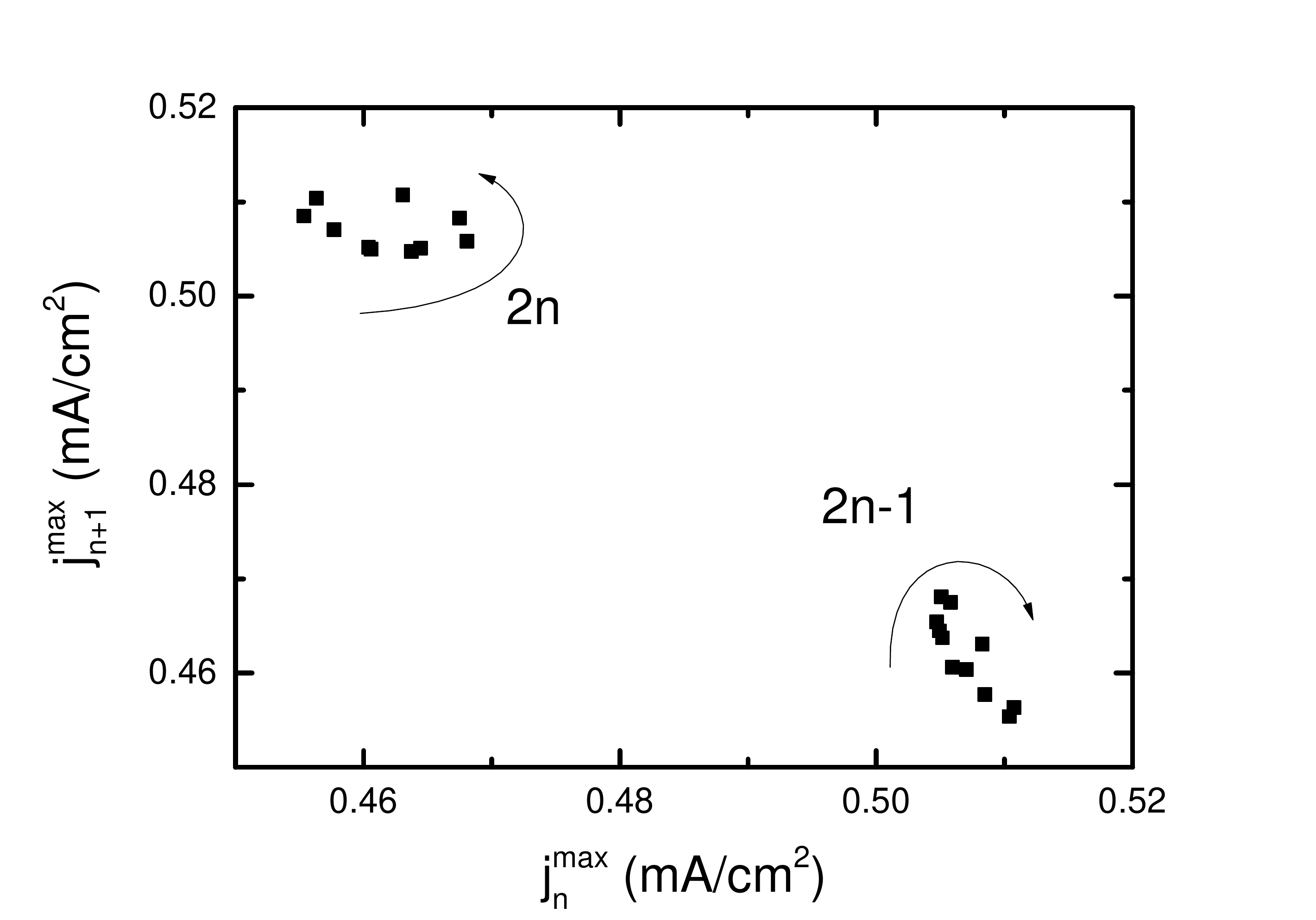}
	\caption{Next-maximum-map of a period-two torus measured at 5.74\,V vs.  SHE; even-numbered points appear in the upper left and odd-numbered points in the lower right corner. The arrows indicate the direction in
		which the points appear.}
	\label{fig:p2tor}
\end{figure}

At the low potential end of the period doubling and halving sequences the dynamics deviates from a classical Feigenbaum scenario. 
Rather than undergoing a transition to a period-1 oscillation, the P2 state (green dots in Fig.\,\ref{fig:feignmm}) gains a further oscillation frequency at $U_\text{app}=5.74$\,V (brown dots in Fig.\,\ref{fig:feignmm}) and thus becomes a period-two torus. 
The next-maximum-map of a P2-torus is depicted in Fig.\,\ref{fig:p2tor} in more detail. 
\newline  \indent 
It should be mentioned that the two circular limit sets seen in Fig.\,\ref{fig:p2tor} tend to spiral towards smaller radii for long measurement times. 
The variations over time might again be due to slight parameter drifts. 
Since the parameter regime in which this period-two torus exists is very small it was so far not possible to prove that this state is stable over a long period of time. 
At even lower voltages the branch of period doubling loses its stability and the system once again attains a state on the branch of the toroidal attractor emerging from the LA oscillations.
\section{Discussion}
\label{sec:dis}
Oscillatory dynamics during the electrodissolution of Si in fluoride-containing electrolytes has been investigated for more than half a century \cite{Turner1958,Zhang2001,Foll2009}.
Yet, the oscillation mechanism still lies in the dark. 
Furthermore, in an attempt to establish identical surface conditions at each parameter, the typical measurement protocol involved to re-establish a H-terminated surface at open circuit potential between parameter variations. 
Abstaining from the established protocol \cite{Miethe2009, Miethe2012, Schoenleber2012,Ozanam1992}, and introducing only slight parameter variations our dynamical study evidenced two types of hystereses. One involves the electropolishing branch and the oscillatory region, the other one emerges from the existence of two intrinsically different limit cycles. 
Each of the two types of limit cycles live on their own branch in a bifurcation diagram, thus being necessarily related to different feedback loops causing the oscillatory instability. In other words, the system is intrinsically birhythmic. Yet, simple birhythmicity was not observed. In contrast, bistability occurred between complex oscillations in a broad parameter band, implying that the two different limit cycles are strongly and bidirectionally coupled. This is different from literature examples, where simple birythmicity could be observed, whereas the interaction between the two oscillations led to complex or compound oscillations but not to a hysteretic behaviour. In tiny parameter regions simple oscillations were reported to coexist with complex ones, but a coexistence of complex oscillations was not detected \cite{Decroly1982,Alamgir1983,Alamgir1984,Johnson1991}.
In our Si system on the other hand, a bistability between two different routes to deterministic chaos could be identified, each of them on one of the two branches.  
This extraordinary feature, which to the best of our knowledge has not yet been observed, makes the system an outstanding model system for birhythmicity, revealing general questions concerning birhythmic dynamics. 
The most intriguing one from our experiments is the question through which bifurcations the two branches undergoing transitions to chaos are linked. 
More specifically, one might ask whether the torus bifurcation of the period-2 oscillation close to the existence boundary of the 'Feigenbaum branch', c.f. Fig.\,\ref{fig:p2tor}, is a necessary step for the annihilation of this branch in some type of saddle node bifurcation, where the saddle-type limit set originates from the 'torus branch'. 
From a broader perspective, the experiments reveal the necessity for a theoretical foundation of bichaoticity. 
Considering that in a broad band of our parameter plane, which separated LA oscillations and HA oscillations, complex dynamics were found, further detailed bifurcation studies might reveal other distinct and possibly novel transition scenarios between branches with chaotic dynamics. 

\section{Conclusion}
\label{sec:conclusion}
Our studies revealed that in order to understand the mechanisms behind the current oscillations in the anodic electrodissolution of silicon one has to differentiate between dynamical phenomena of different origin. 
On the one hand, we observed a bistability between a stable node and oscillatory branches, which expands the hitherto known parameter space for stable oscillations drastically. This bistable regime is enclosed by two saddle node bifurcations.
On the other hand, we demonstrated that one should distinguish between three types of oscillatory dynamics, simple periodic LA oscillations, simple periodic HA oscillations and complex oscillations. We presented evidence that the HA and LA oscillations arise through different feedback loops. The interaction between these are so strong that in a large parameter region both of them undergo a transition to chaos, the LA oscillations through a quasiperiodic scenario and the HA oscillations through a period doubling sequence. Through which bifurcations the branches are connected remains an open question and points to the necessity to further investigate bichaoticity both theoretically and experimentally.
\begin{acknowledgments}
KK would like to thank Ken Showalter for many inspiring discussions throughout
her academic life. The project was funded by the Deutsche
Forschungsgemeinschaft (DFG, German Research Foundation)
project KR1189/18.
\end{acknowledgments}
\section*{References}
\bibliography{bibliography}
\end{document}